\documentstyle[11pt,paspconf]{article}

\begin{document}

\title{Dynamics of Galaxies in Clusters}

\author{Oleg Y. Gnedin}

\affil{Princeton University Observatory}

\section{Motivation: Spirals versus S0s}

Recent observations by the Hubble Space Telescope show that spirals make
up to 50\% of the galaxies in clusters at redshift $z \sim 0.5$
(Dressler et al. 1997; Ellis 1998).  Spectroscopic evidence suggests
bursts of recent, or even current, star formation.  This picture is
drastically different from the nearby clusters, populated mostly by
featureless elliptical or S0 galaxies.  The change is due to the
evolution of S0s, which are depleted by about a factor of two at $z \sim
0.5$.  Observations suggest that star formation has ceased in the
ellipticals by an epoch $z \sim 0.6$, but continued in the S0s until at
least $z \sim 0.3$ (van Dokkum et al. 1998).  It is apparent that a
strong dynamical and morphological evolution of disk galaxies has
occurred in clusters of galaxies.

\section{Self-Consistent Cluster and Galactic Simulations}

In order to study the self-consistent evolution of galaxies in the
cosmological context, I use a large Particle-Mesh numerical simulation.
The PM code is coupled with the Self-Consistent Field code to study the
dynamics of individual galaxies with the very high effective resolution.

A cluster of galaxies is simulated using constrained Gaussian initial
conditions in each of the three cosmological scenarios ($\Omega_0=1$;
$\Omega_0=0.4$; and $\Omega_0=0.4$, $\Omega_\Lambda=0.6$).  The
simulation box has the comoving size of 32 $h^{-1}$ Mpc and $512^3$ grid
cells.  In these simulations, I follow the trajectories of the
identified galaxies and calculate the external tidal field along them.
Then, I use the tidal histories of the galaxies as an input to the
galactic SCF simulations.  Each galaxy is represented by $10^6$ star
particles and $10^6$ dark matter particles, with the spherical halo
being 20 times more massive than the stellar disk.  I carefully document
and, whenever possible, subtract the effects of numerical relaxation.

\section{Results}

In agreement with analytical models, the dark halos of the massive
galaxies are truncated at 30 to 60 kpc, regardless of their initial
extent.  As a result, the galaxies lose a significant fraction of their
mass.  Inside the truncation radius, the halo density profile is
perturbed very little.  The stellar disk maintains most of its mass and
the initially exponential surface density stays about the same.
However, there is a substantial amount of heating in the vertical
direction: the scale-height increases by a factor of 1.5 to 2.5,
depending on the tidal history.  Toomre's $Q$ parameter increases to $Q
\sim 2$, making the disk gravitationally very stable.  In cooperation
with the stripping of gas by the intracluster medium, this effect halts
new star formation in the galaxies.  The end result is the morphological
transformation of spiral galaxies into the S0s.

Figure 1 shows an example of a large spiral galaxy at the end of the
simulation.  The visible extent of the galaxy is five radial
scale-lengths, or about 15 kpc.  The large outer spiral arms are caused
by tidal perturbations, but any gas is likely to be stripped from the
outer regions.  Looking edge-on, the disk is thick and warps in the
outer parts.  It looks very similar to the image of an S0 galaxy NGC
4762 (Sandage 1961).

Overall, the predicted evolution of galaxies is more modest than in the
similar study by Moore et al. (1996).  The disks of large galaxies
survive but thicken and cease star formation.  The main cause of the
difference is the time evolution of the tidal force during cluster
formation.  The low $\Omega_0$ models agree with observations better
than the $\Omega_0=1$ model.  Full results of this study will be
published elsewhere (Gnedin 1999).

\begin{figure}
\vspace{2in}
\includegraphics{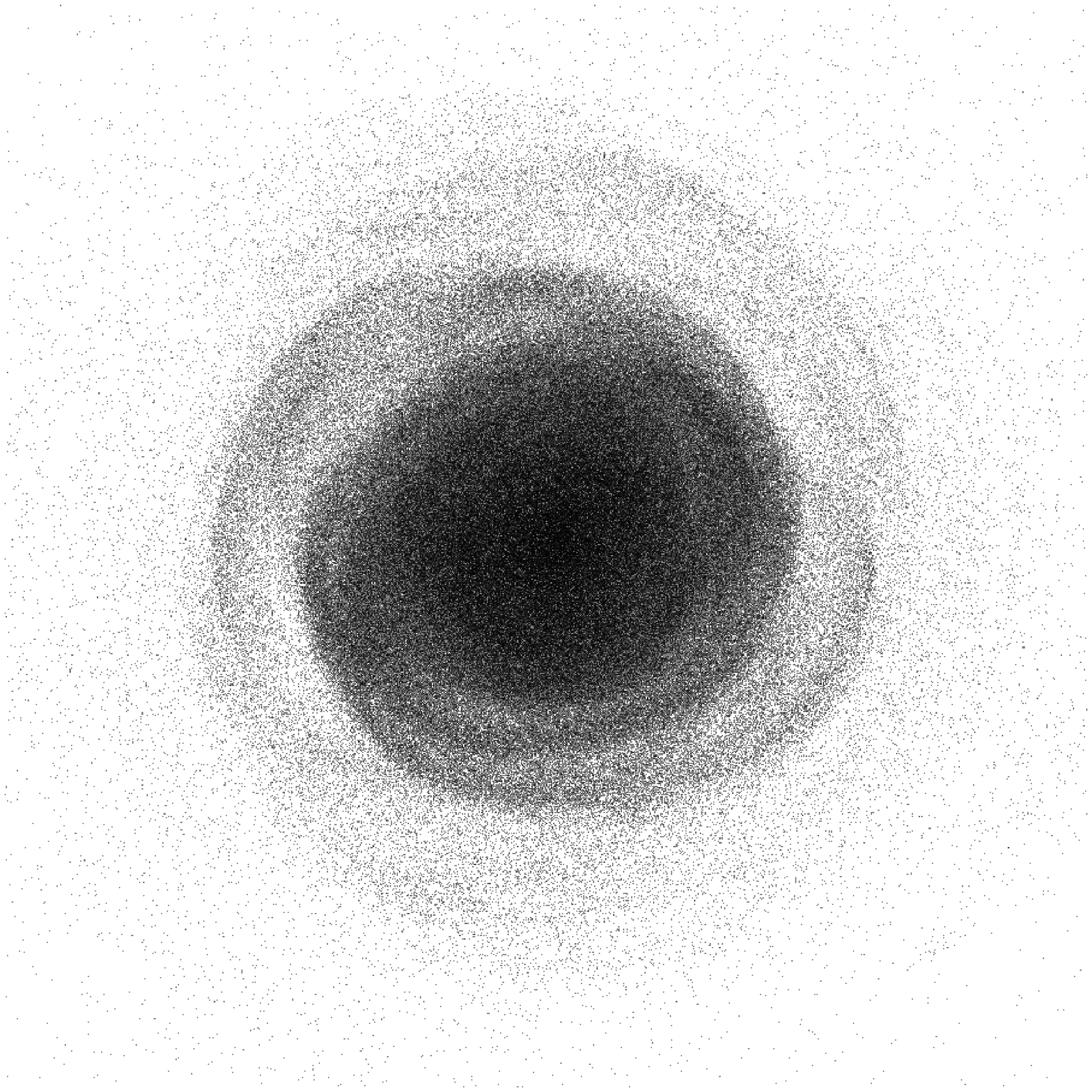}
\includegraphics{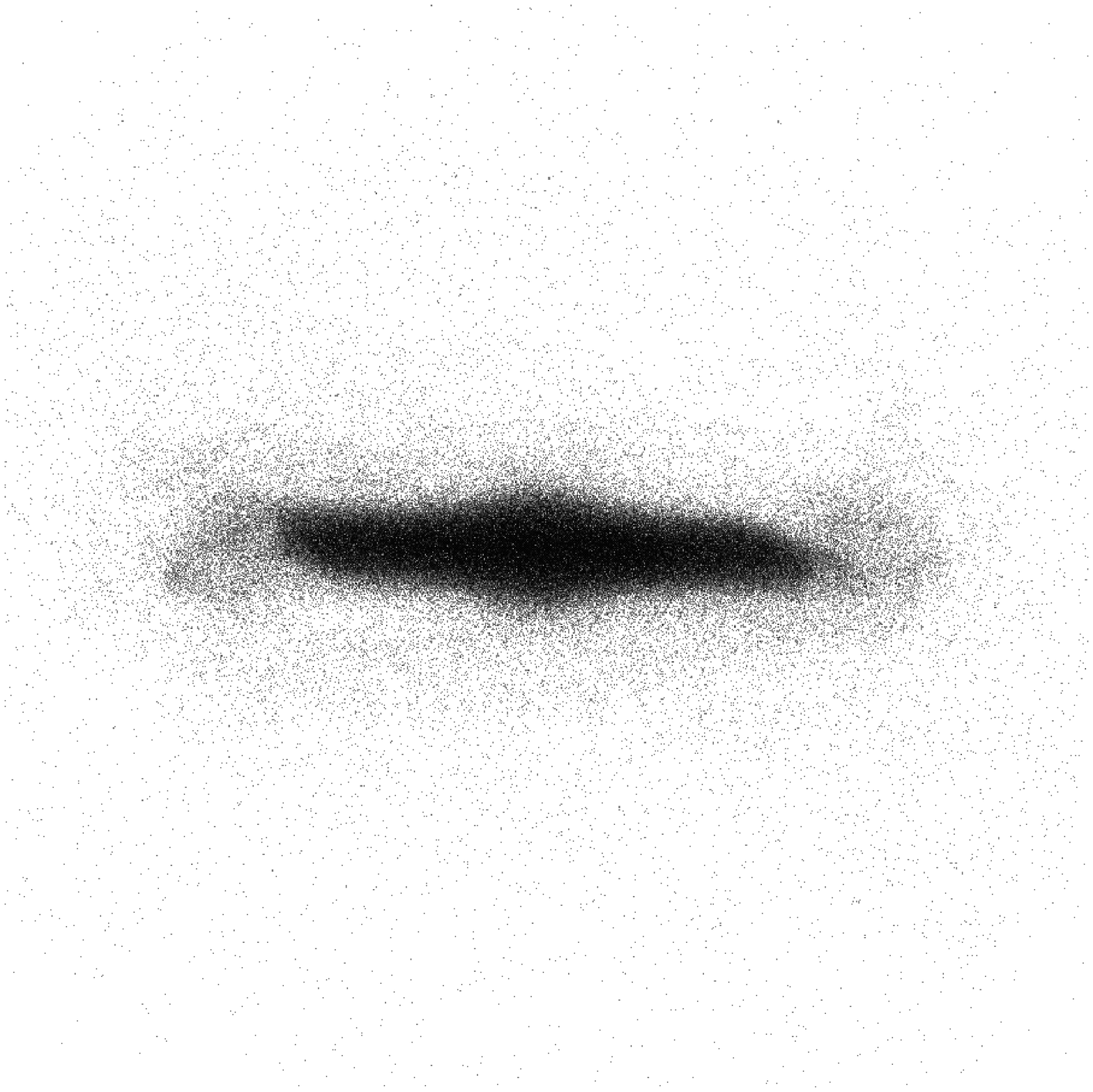}
\caption{Face-on and edge-on views of the final configuration of an initially
 large spiral galaxy.}
\end{figure}

\acknowledgments

I wish to thank Lars Hernquist and Renyue Cen for letting me use their
SCF and PM codes, respectively.  I am grateful to Jerry Ostriker for his
continuous support and guidance.


\begin{references}

\reference
Dressler, A., et al. 1997, ApJ, 490, 577

\reference
Ellis, R. 1998, Nature, 395 (Suppl.), A3

\reference
Gnedin, O. Y. 1999, ApJ, submitted

\reference
Moore, B., Katz, N., Lake, G., Dressler, A., \& Oemler, A. 1996,
  Nature, 379, 613

\reference
Sandage, A. 1961, The Hubble Atlas of Galaxies (Washington,DC: Carnegie
  Institution of Washington)

\reference
van Dokkum, P. G., et al. 1998, ApJ, 500, 714

\end{references}
\end{document}